\documentclass[10pt,conference,letterpaper]{IEEEtran}
\pdfoutput=1

\usepackage{cite}
\usepackage{amsmath,amssymb,amsfonts}
\usepackage{url}

\usepackage{amsthm}
\theoremstyle{definition}
\newtheorem{definition}{Definition}[]
\newtheorem{theorem}{Theorem}[]

\usepackage[noend]{algpseudocode}
\usepackage{graphicx}
\usepackage{textcomp}
\usepackage{xcolor}
\usepackage{enumerate}

\usepackage[font=small]{caption}
\usepackage{subcaption}

\newcommand{\our}{\textit{CollabFed}}

\usepackage[ruled,vlined, linesnumbered]{algorithm2e}

\def\BibTeX{{\rm B\kern-.05em{\sc i\kern-.025em b}\kern-.08em
    T\kern-.1667em\lower.7ex\hbox{E}\kern-.125emX}}
    
\begin{document}
\title{Leveraging Public-Private Blockchain Interoperability for Closed Consortium Interfacing}

\author{\IEEEauthorblockN{Bishakh Chandra Ghosh, Tanay Bhartia,
		Sourav Kanti Addya, and Sandip Chakraborty}
	\IEEEauthorblockA{Department of Computer Science and Engineering\\ Indian Institute of Technology Kharagpur, India\\
		Email: ghoshbishakh@gmail.com,
		tanaybhartia@gmail.com,
		souravkaddya@cse.iitkgp.ac.in,
		sandipc@cse.iitkgp.ac.in}}

\maketitle

\begin{abstract}
With the increasing adoption of private blockchain platforms, consortia operating in various sectors such as trade, finance, logistics, etc., are becoming common. Despite having the benefits of a completely decentralized architecture which supports transparency and distributed control, existing private blockchains limit the data, assets, and processes within its closed boundary, which restricts secure and verifiable service provisioning to the end-consumers. Thus, platforms such as e-commerce with multiple sellers or cloud federation with a collection of cloud service providers cannot be decentralized with the existing blockchain platforms. This paper proposes a decentralized gateway architecture interfacing private blockchain with end-users by leveraging the unique combination of public and private blockchain platforms through interoperation. Through the use case of decentralized cloud federations, we have demonstrated the viability of the solution. Our testbed implementation with Ethereum and Hyperledger Fabric, with three service providers, shows that such consortium can operate within an acceptable response latency while scaling up to 64 parallel requests per second for cloud infrastructure provisioning. Further analysis over the Mininet emulation platform indicates that the platform can scale well with minimal impact over the latency as the number of participating service providers increases.
\end{abstract}



\section{Introduction}\label{intro}
Business-to-Business (B2B) and Business-to-Consumer (B2C) online marketplaces have gained much attention nowadays within various sectors, including e-commerce, ride-hailing, cloud service provisioning (e.g., cloud federations), supply-chain management, etc.. However, there has been a continuing debate about the market-monopoly and unfairness they created in the digital economy~\cite{haucap2014google,hindman2018internet}. Such platforms typically work as the central agent or broker to interconnect various businesses and consumers. In such a firm-controlled marketplace, supporting trustworthiness and unbiased business transactions is always a concern. Blockchain is a natural extension to help trustworthy and bias-free business by allowing the stakeholders to interact over a decentralized marketplace. Therefore, various recent works advocate for developing blockchain-based electronic marketplaces~\cite{subramanian2017decentralized,hynes2018demonstration,blockv,artchain,chang2019blockchain,narang2019design}. However, there is a fundamental limitation of the current blockchain technologies to support this, as discussed next. 

An electronic marketplace is typically a multifaceted network with one or more closed business networks collaborating through B2B transactions and, finally, an open consumer network having B2C operations~\cite{schiff2003open}. For example, in a typical supply chain, manufacturers, wholesalers, and retailers form different closed business networks, and finally, the end-customers create an open consumer network. Another example is cloud federation platforms like OnApp\cite{onapp}, where small cloud service providers (CSPs) construct a closed consortium to provide cloud resources to customers. Depending on customer requests, the transactions flow from the open consumer network to various closed business networks, and the service is finally delivered back to the open consumer network. 

Emerging blockchain networks such as \textit{IBM Food Trust}\cite{ibmfoodtrust}, \textit{TradeLens}\cite{tradelens}, \textit{Marcopolo}\cite{marcopolo}, etc., use private (permissioned) distributed ledger-based systems like \textit{Hyperledger Fabric}~\cite{fabric} and \textit{Corda}~\cite{corda} to form closed consortia of businesses. However, a key limitation of the existing private blockchain platforms is the restriction of their applicability within only closed consortia where data and assets are not required to be communicated outside the network boundary. Thus, \textit{Fabric}, \textit{Corda}, or other existing private blockchains do not support any interface or protocols for interacting with the open network outside, which is crucial for building consortia of service providers acting together to deliver services to the consumer network. 

However, there are challenges in designing such interfacing. \textbf{First}, the businesses, as well as the consumers, can exhibit byzantine behavior in the absence of a firm-controlled marketplace. 
Therefore they can collude to deceive and take control over the consortium decisions.
\textbf{Second}, the consumers' service requests need to be agreed upon by the businesses within the closed consortium along with their ordering, before they can be processed. Otherwise, any malicious business can take priority over a profitable service request, thus affecting the fairness of the system. Although private blockchain can ensure transaction execution order within the closed network, they do not support transactions from outside the closed network pertaining to Sybil attacks from the open network participants~\cite{sybilattack}. \textbf{Third}, the service responses from the closed consortium also need to be transferred back to the consumer who requested the service. Such information must be verifiable by the consumers against the valid consensus at the business network. Further, the privacy of the information must be ensured.

Thus, towards developing a decentralized collaborative architecture for service providing consortia, we introduce {\textbf{\our}}, which addresses the above challenges by building a novel decentralized interface between the private blockchain networks and the open network of consumers. {\our} ensures multi-party consensus validation and considers threats such as Sybil attacks and byzantine behaviors of the participants. The decentralized interface is engineered through a unique combination of the public blockchain and private blockchain networks by enabling interoperability between them to support trusted and secure data transfer in both the directions, that is (a) from the consumers to the businesses and (b) from the businesses to the consumers (\textbf{Contribution-1}). Our \textit{Consensus on Consensus} mechanism handles the transfer of data from the open network into the private blockchain in a secured and verifiable manner (\textbf{Contribution-2}). We employ a novel mechanism based on \textit{collective signing} (CoSi) technology~\cite{syta2016keeping} to generate verifiable results from the consortium, which is accessed securely by the consumers (\textbf{Contribution-3}). Moreover, {\our} facilitates the collaboration among the participating businesses and enables fair scheduling of requests through a distributed consensus. Performance in terms of latency is of utmost importance here, so we analyze the effect of \textit{order-execute} and \textit{execute-order} transaction execution workflows on the performance of request scheduling.

Considering a use case of a decentralized brokerless cloud federation, we have done a proof-of-concept (PoC) implementation of {\our} using \textit{Ethereum} as the public blockchain platform and \textit{Hyperledger Fabric}, and \textit{Burrow} as the two different candidates for the private blockchain platform, and tested it with three emulated CSPs (\textbf{Contribution-4}). The experiments prove the viability of {\our} as a platform for service provisioning consortia, which supports interaction between a private blockchain network and the end-consumers. Evaluation of the performance shows acceptable overhead on the federation, and a Mininet-based emulation with $32$ CSPs also validates its scalability over a large geo-distributed setup.

\section{Related Work}\label{relatedWork}
One of the most compelling use cases of blockchain technology is in industries and enterprise environments where multiple authoritative domains such as companies, organizations, and governments form a consortium without any central trusted mediator's involvement. Research on enabling such applications have been carried out in sectors like energy trading~\cite{gai2019privacy}, supply chain~\cite{supplychain}, cloud~\cite{hu2018searching,zhou2019blockchain, ourscc}, 
and many more~\cite{belotti2019vademecum}. However, almost all existing solutions consider a closed consortium of organizations that do not require communication with the outside. Some blockchain-driven systems which enable businesses to interact with consumers have been proposed, such as BlockV~\cite{blockv}, a ride-sharing application ensuring fairness, and ArtChain~\cite{artchain} - a blockchain-based art marketplace. Similarly, Savi \textit{et al.} introduced a public blockchain-based cloud brokerage platform~\cite{blockchainbroker} using Ethereum for sharing spare fog resources. Although connecting businesses and consumers, these platforms are based on the public blockchain only, and thus are not suitable for enterprise use cases that involve sensitive data exchange between the consortium members. Moreover, public blockchains are not ideal for complex business logic-based smart contracts since they have to be replicated and executed over the entire network, thus hampering performance.

Using private blockchain for such use cases 
will require some mode of interoperability with the public blockchain. Several prior works focus on cross-chain communication~\cite{sok} for different applications such as cross-network asset exchange or asset transfer. Most of them such as, Tesseract~\cite{tesseract}, Herlihy~\cite{herlihy2018atomic}, Xclaim~\cite{xclaim}, AMHL~\cite{malavolta2019anonymous}, 
 focus on public-public blockchain interoperability for exchange of cryptocurrency, asset transfer, and payment channel networks. On the other hand, Omniledger's Atomix~\cite{omniledger}, Chainspace~\cite{chainspace}, Fabric Channels~\cite{fabricchannels} enable interoperability and transactions between different shards of the same blockchain platform. Abebe \textit{et al.}~\cite{abebe2019enabling} proposed a protocol for trusted and verifiable data transfer across private blockchain networks using endorsement collection. Cash \textit{et al.}~\cite{cash2018two} proposed a two-tier public-private blockchain architecture for secure data sharing. However, none of these existing works address the interoperability and data transfer between private and public blockchain platforms.

To the best of our knowledge, {\our} is the first attempt to address the issue of communication of consumer requests, and processed responses between a private blockchain-based consortium and the open network through public-private blockchain interoperability.
\section{System Model and Design Challenges}
We consider the interconnecting network between the consumers and the closed consortium to be partially synchronous where there is an upper bound $\Delta$ on the time of message delivery~\cite{liu2016xft,dwork1988consensus}. If a message is not received within the time-bound $\Delta$, then it is considered as a message fault. The intuition is that in a realistic communication, the messages must have arbitrary but bounded delay. This results in challenges such as unordered message delivery and message drops. Additionally, we consider different types of attacks that might affect the above operations, as follows.

\subsection{Threat Model}
A decentralized consortium is prone to the following types of attacks, which we take care of in the design of \our{}. 

\noindent\textbf{Byzantine participants:}
We consider that at most $\frac{1}{3}$ of the participants, both for businesses and consumers, may exhibit byzantine behavior~\cite{lamportbft,pbft,aublin2013rbft,bftsmart}. A consumer can try to deceive the consortium by sending different requests to different businesses, while the businesses can collude themselves to alter the decision protocols' results to take control of the consortium. 

\noindent\textbf{Sybil attacks:}
BFT consensus protocols assume that each participant has only one distinct identity~\cite{pbft, aublin2013rbft, dwork1988consensus}. If somehow one participant can generate multiple identities, then using such redundancy, it can launch a ``Sybil Attack"~\cite{sybilattack}. The consumers thus can launch a Sybil attack to the closed consortia by using multiple identities. 

\noindent\textbf{Impersonation attacks:}
As a decentralized architecture, the consortium does not have a single spokesperson responsible for communicating with the open network consumers. Exploiting this, a malicious business from the closed consortium might try to deceive a consumer by posing as the consortium's spokesperson and providing false information. 

\noindent\textbf{Leakage of sensitive information:}
The business and the consumers communicate over an open, unsecured channel through message passing. Therefore, sensitive information like credentials, contact information, etc., might get leaked.

\subsection{Design Philosophy and Challenges}
\our{}'s primary objective is to develop a mechanism through which any closed consortium designed using a private blockchain platform can interface with open consumer networks. Considering the threat model as discussed above and the possibility of unordered message delivery along with message drops, in \our{}, the following two guarantees need to be ensured at the consortium interface. 
\theoremstyle{definition}
\begin{definition}{\textbf{Consortium Interface Safety -}}
	The interface should ensure that all the correct consortium members agree on the same set of incoming consumer requests in same order. 
\end{definition}
\begin{definition}{\textbf{Consortium Interface Liveness -}}
	The interface must ensure that all the correct consumer requests are eventually be processed and committed by the closed consortium.
\end{definition}

Thus, a mechanism is needed such that the interface meets the safety and liveness guarantees, and the consortium members are in a consensus on each request. To achieve consensus over the ordering of consumer requests from the open network, we propose to use
public blockchain platforms~\cite{ethereum, algorand, proofofstake} for interfacing the closed consortium to the consumers of the open network. The consensus algorithms over a public blockchain setting are designed to be resistant to Sybil attacks. Therefore, using a public blockchain platform, the consumers' requests from an open network can be ordered. However, merely clubbing together any public and private blockchain is not enough to enable the targeted consortium interface; there are open challenges that need to be solved. 
\begin{enumerate}[(i)]
\item \textbf{Passing consensus of one network to another:} The public and the private blockchain networks run their own consensus protocols independently. The interface should pass the consensus information from one network to another by ensuring (i) security, and (ii) accountability. The interface should guarantee that the consensus information of one network is verifiable at the other network.  
\item \textbf{Transferring sensitive information from the closed network to the consumers:} Once a consumer request is scheduled and processed by the closed consortia, the associated service information such as access credentials, invoice, shipping information, etc., need to be passed to only the targeted consumer who has requested for the service. Therefore, merely putting the information to the public blockchain will not help, as anyone will access it. Protocols need to be designed to share such sensitive information with the targeted consumer only. 
\item \textbf{Verifiability of the consortium decision:} The consortium's decision of scheduling, service provisioning, etc. comes through a consensus over the private network. However, once this information is forwarded to the public network, the consumers should be able to verify such decisions to avoid any byzantine behavior from the colluded consortium members.
\end{enumerate}

\section{Decentralized Consortium Interface}
\begin{figure}[!ht]
	\centering 
	\includegraphics[width=0.9\linewidth]{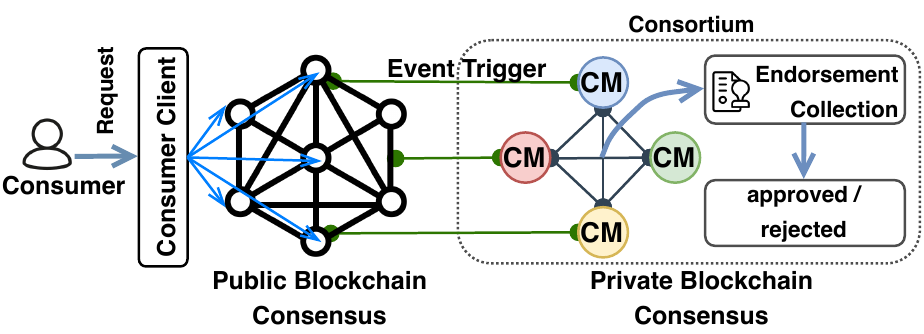}
	\caption{Transferring Consumer Requests from Public Blockchain to the Consortium Members (CMs)}
	\label{fig:usertosp}
\end{figure}

\begin{figure}[!ht]
	\centering 
	\includegraphics[width=0.9\linewidth]{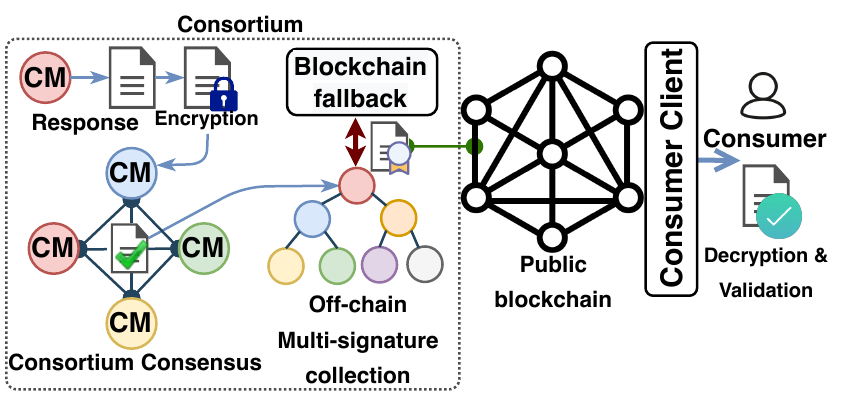}
	\caption{Secure and Verifiable Data Transfer from CMs to Consumers}
	\label{fig:sptouser}
\end{figure}

The functionality of \textit{{\our} Consortium Interface} is broadly two-fold:
(a) transferring consumer requests from the open network to the closed consortium members (Fig \ref{fig:usertosp}), (b) transferring consortium responses to the open network consumers in a secure and verifiable way (Fig \ref{fig:sptouser}). The \textit{Consortium Interface Safety} is achieved using two rounds of consensus over the consumer requests -- (1) \textit{regular consensus (mining) of the public blockchain}, and (2) A \textit{Consensus on Consensus} mechanism. The details follow. 

\subsection{Regular Consensus (Mining) over Public Blockchain}
Before scheduling and processing any consumer request, the consortium members must reach a consensus on the same. Moreover, there needs to be a consensus on the order in which the requests are to be considered to ensure \textit{Consortium Interface Safety}.
This ensures that a malicious member of the consortium cannot collude the network by triggering the scheduling of an invalid consumer request or take priority on a specific consumer request.
\our{} uses public blockchain in conjunction with the private consortium to support this. However, for supporting interoperability between the two networks, the consensus has to be propagated between them.

To interact with the consortium, consumers send their requests through the public blockchain. These requests are formed as transactions to a smart contract - \textit{User Request Contract}, deployed in the public blockchain. Just like a web interface of a central firm-controlled platform, this smart contract acts as the communicating point for the consumers to reach the consortium, albeit in a decentralized way. The ``consumer request" transactions are then committed to a block in the ledger through the public blockchain platform's mining/consensus process. For example, \textit{Ethereum} uses a modification of the most popular consensus protocol: ``proof of work" (PoW)~\cite{bitcoin}, while there  are many alternate consensus protocols, such as Proof of Stake~\cite{proofofstake}, Bitcoin-NG~\cite{bitcoin-ng}, Byzcoin~\cite{byzcoin}, Algorand~\cite{algorand} etc. These consensus protocols have different safety and liveliness assumptions of their own; however, their common objective is to reach consensus on a block of transactions. Moreover, since these are permissionless blockchain protocols, they are designed to resist Sybil attacks.

Once a block is mined and committed in the public blockchain, this ensures that there is a consensus on the particular block and their order in which they are committed, since each block is linked to the previous one through its cryptographic hash. Moreover, the set of transactions in each block also has a fixed packing order for the smart contracts' deterministic serial execution. Despite these properties, public blockchain consensus itself is not enough to satisfy \textit{Consortium Interface Safety}, and consortium members cannot simply pick user requests from the public blockchain and start processing them. The reasons are as follows. (1) Due to the partially synchronous network, some consortium members might not get the mined block in time and thus cannot participate in its scheduling. (2) Malicious consortium members may introduce and schedule invalid consumer requests that are not mined at all. (3) Public blockchain consensus protocol like PoW, often goes through temporary forks~\cite{forkanalysis}, resulting in conflicting consumer requests or conflicting ordering in different members. Thus, \our{} has to carry out a second round of consensus, which we call \textit{Consensus on Consensus}.

\subsection{Consensus on Consensus}
In~\cite{sok}, the authors have shown an interesting result that states that cross-chain communication is impossible without a trusted third party. To circumvent this impossibility result, {\our} uses a novel idea where the private consortium members also participate in the public blockchain to represent themselves as their own trusted agent. Whenever a new block is committed in the public blockchain, the trusted agents corresponding to the private consortium members get an event-trigger, which in turn invokes a \textit{Propagation Contract} in the private blockchain network. Before invoking the \textit{Propagation Contract}, the transactions of the public blockchain can be verified individually by the consortium members by existing methods such as \textit{Simplified Payment Verification (SPV)} as used in standard public blockchain like Bitcoin~\cite{bitcoin}.  

\begin{figure}[!ht]
	\centering 
	\includegraphics[width=0.95\linewidth]{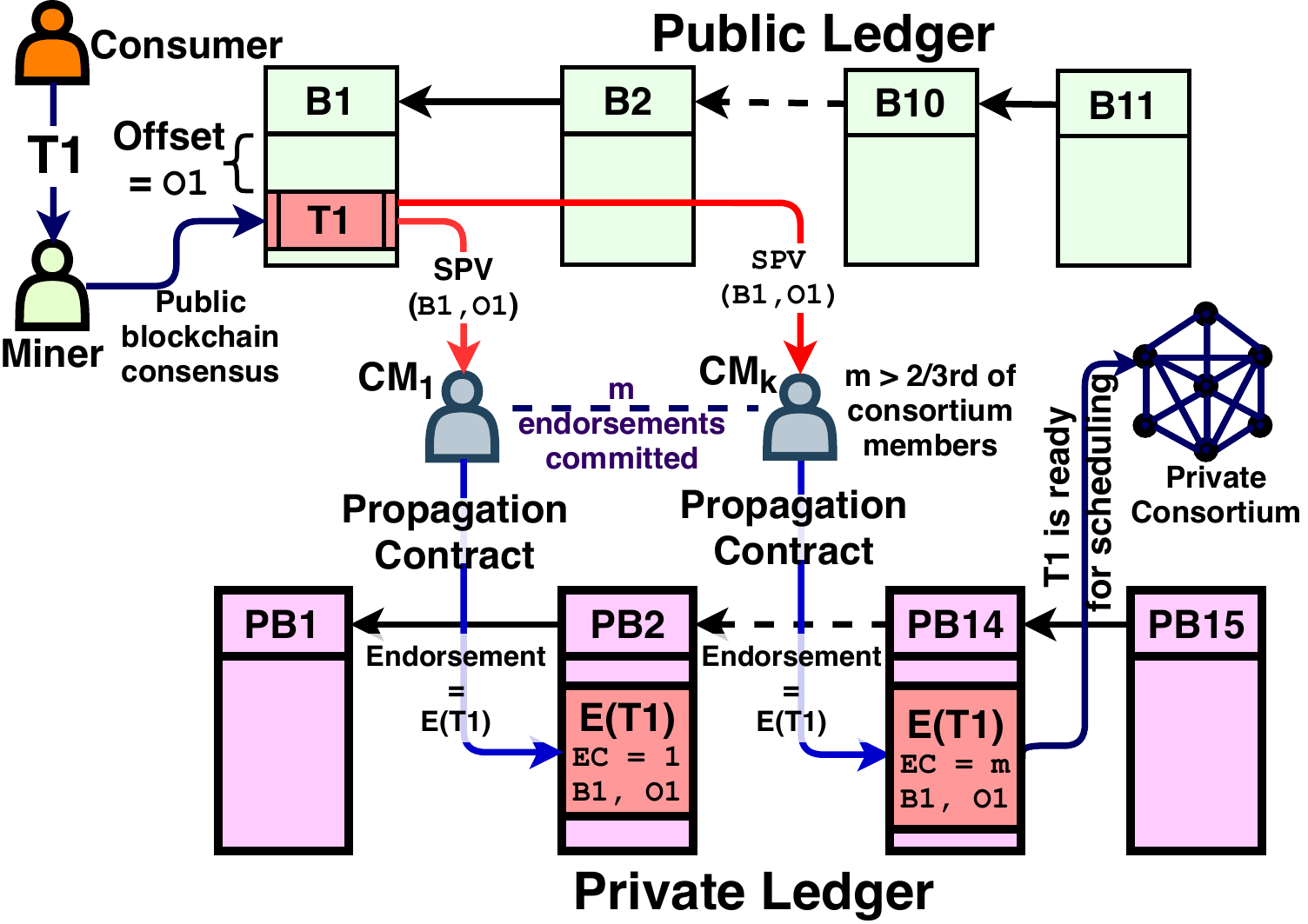}
	\caption{Propagation Contract: \textit{Consensus on Consensus}}
	\label{fig:consensuspropagation}
\end{figure}

The task of the \textit{Propagation Contract} is to collect \textbf{verification endorsements} from consortium members for each consumer request. The \textit{verification endorsements} are the digitally signed certificates from the consortium members, indicating that the corresponding members agree on the processing of a consumer request committed over the public blockchain. As per the standard BFT protocols~\cite{pbft,byzcoin}, a consumer request can be committed for scheduling in the private consortium if the majority ($\frac{2}{3}$rd) of the consortium members endorse the request transaction. The endorsement protocol used in the \textit{Propagation Contract} is shown in Fig.~\ref{fig:consensuspropagation}. The details follow.  

\noindent\textbf{Endorsement Initialization:} Whenever a consortium member receives a ``consumer request'' transaction through the event listener of the public blockchain, it checks whether there is already an endorsement available in the private ledger corresponds to that transaction. If no endorsement is available, it initiates the endorsement collection process for that particular request by initiating the \texttt{endorsement-count (EC)} variable set to $1$, and committing the signed endorsement in the private ledger. The request is also accompanied by a sequence number for representing its order. This sequence number is formed as \texttt{\{blocknumber, offset\}}, indicating the block in which the request transaction is committed in the public blockchain, and its packing order inside the block. 

\noindent\textbf{Endorsement Propagation:} As other consortium members also get the same consumer request and with the same \texttt{\{blocknumber, offset\}} through the event listener of the public blockchain, they also execute the \textit{Propagation Contract} for it, which adds their signed endorsements while incrementing the \texttt{EC}. Each execution of the \textit{Propagation Contract} is also a transaction. Therefore, each endorsement also goes through the consensus process of the private blockchain. 

\noindent\textbf{Commitment:} Thus, the number of endorsements for a request goes up until it reaches greater than two-third of the number of consortium members ($\texttt{EC} > \frac{2}{3} |consortium|$). At this point, the majority of the consortium participants have consensus on the request through endorsements, and each such endorsement has a consensus of the network. Thus, the consumer request is marked as approved and ready to be scheduled. 

\begin{theorem}
The \textit{Consensus on Consensus} mechanism ensures consortium interface safety and consortium interface liveness. 
\end{theorem}
\begin{IEEEproof}
Whenever a transaction is committed in a block in the public blockchain, it implies all its correct participants including consortium members agree on it, along with the (\texttt{blocknumber}, \texttt{offset}).
The \textit{Consensus on Consensus} mechanism endorses the transactions from the public blockchain and then commits the endorsements in the private blockchain. A transaction is scheduled only when more than $\frac{2}{3}$ of the consortium members endorse the transaction.
Given that each endorsement transaction also undergoes consensus in the private blockchain, and the given verifiability property of the private ledger, a transaction from the public blockchain is executed only when the majority of the consortium members endorse it. Further, the transactions are executed in the order of (\texttt{blocknumber}, \texttt{offset}) parameters of the public blockchain ensuring agreement on the order.
Thus the \textit{Consensus on Consensus} mechanism ensures interface safety. 

Consortium interface liveness depends on the liveness of the public blockchain. The event-listeners for correct consortium members eventually trigger the propagation contract when a transaction is committed in the public ledger. Even if there is a temporary fork, the propagation contract is executed when the transaction is finally committed in the public ledger. 
\end{IEEEproof}

The \textit{Propagation Contract} triggers \textit{Scheduling Contract} that schedules the requests based on a predefined business logic. After a request is scheduled and processed over the closed consortium, the service results have to be transferred back to the consumers. The details follow. 

\subsection{Secure and Verifiable Response Transfer}
\label{subsec:secureinfotransfer}
A consortium is operated collectively by its participant businesses. Hence, any data/information provided by it has to be the result of the collective consensus process. Thus, in the absence of a central coordinating platform, this consensus has to be collected and verified by the consumers, without depending on any trusted agent. There can be two variations of information originating from the consortium. (1) \textbf{\textit{Consortium information}} such as information about the participating businesses, service catalogs, etc., and (2) \textbf{\textit{Request responses}} that are the results of scheduling and processing consumer requests such as a digital document.

Both of these kinds of data are generated collectively by the consortium members through the private blockchain's consensus process. However, this consensus information has no manifestation outside this closed network. Thus, consumers being outside the consortium and not participating in the consensus protocol cannot verify the correctness of the data that is committed through transactions in the private blockchain. A separate protocol has to be designed through which information transfer from the consortium to the consumers can be validated outside the private network concerning the consensus of the participating businesses. Moreover, although \textit{consortium information} can be considered publicly available, the \textit{Request responses} to the consumers may contain sensitive information that should remain confidential while being transferred across the open network of the public blockchain.

In \our{}, we use the concept of \textit{Collective Signing} (CoSi)~\cite{syta2016keeping} where a set of consortium members collectively sign a valid information to make it verifiable. We utilize \textit{Boneh-Lynn-Shacham} (BLS) cryptosystem~\cite{bls} for collecting and aggregating signatures from the individual participating businesses. Similar to Byzcoin~\cite{byzcoin}, which uses CoSi to reach to a BFT consensus, a piece of information posted by the consortium through the public blockchain is considered to be valid, if and only if it has been signed by at least $\tfrac{2}{3}$rd of the consortium members. The details follow. 

\subsubsection{BLS Signatures}
A BLS signature is computed as $\mathbb{S}_{i}(\mathcal{M}) = \mathcal{H}(\mathcal{M})^{\mathcal{S}_{\mathcal{C}_{i}}}$, where $\mathcal{M}$ is the message that is to be signed, $\mathcal{H}(.)$ is a cryptographic hash function, and $\mathcal{S}_{\mathcal{C}_{i}}$ is the secret key of the consortium member $\mathcal{C}_{i}$. The property that makes BLS signatures special is that they can readily be extended to multi-signatures. Therefore, for $n$ members participating in the consortium,  $\mathcal{C}_{1}, \mathcal{C}_{2}, \cdot, \mathcal{C}_{n}$, the aggregated multi-signature can be calculated as follows. 
\begin{equation}
\small
\begin{split}
\mathbb{S}_{1..n}(\mathcal{M}) = \mathcal{H}(\mathcal{M})^{\mathcal{S}_{\mathcal{C}_{1}} + \mathcal{S}_{\mathcal{C}_{2}} + .. + \mathcal{S}_{\mathcal{C}_{n}}} = \prod_{i = 1}^{n} \mathcal{H}(\mathcal{M})^{\mathcal{S}_{\mathcal{C}_{i}}}
\\
= \mathbb{S}_{1}(\mathcal{M}) \times \mathbb{S}_{2}(\mathcal{M}) \times . . \times \mathbb{S}_{n}(\mathcal{M}) =  \prod_{i = 1}^{n} \mathbb{S}_{i}(\mathcal{M})
\end{split} \label{eq:bls}
\end{equation}

This aggregated multi-signature $\mathbb{S}_{1..n}(\mathcal{M})$ can be verified with the help of the public keys of the individual consortium members. This verification is done by comparing the pairing operation between the aggregated signatures and the aggregated public keys. The aggregated public key for $n$ members is calculated as $\prod_{i = 1}^{n} \mathcal{P}_{\mathcal{C}_{i}} $, where $\mathcal{P}_{\mathcal{C}_{i}}$ is the public key of $\mathcal{C}_{i}$.

\subsubsection{Posting information using BLS}
Any information about the consortium is communicated to the consumers by posting the same in the public blockchain. Such information originates from the result of the \textit{Collaboration Contract} in the private blockchain, which is responsible for reaching consensus on them. This resultant data like updated information or updated catalog, etc. must be collectively signed by at least $\tfrac{2}{3}$rd of the participating consortium members. This again has two different levels of security requirements for \textit{Consortium information} and \textit{Request responses}. 

\textbf{Posting Consortium Information to the Public Blockchain:} Let $\mathcal{I}$ be a piece of public consortium information that is meant to be seen by all consumers. $\mathcal{I}$ is proposed by a consortium member in the private blockchain where consensus is reached over it. To post this information over the public blockchain, the consortium members over the closed network construct a \textit{Signing-Request message} as $\texttt{sign}\lbrace\mathcal{H}(\mathcal{I}), \mathbb{B},[\mathcal{H}(\mathcal{I})]_{\mathbb{S}_{\mathbb{B}}}\rbrace$ and forward it to all other consortium members. Here $\mathbb{B}$ is a bitmap indicating which members have signed the message and $[\mathcal{H}(\mathcal{I})]_{\mathbb{S}_{\mathbb{B}}}$ is the aggregated collective signature on the hash of the message $\mathcal{I}$. Every consortium member, upon receiving this message, adds its own signature through multiplication, as shown in Eq.~(\ref{eq:bls}), updates $\mathbb{B}$ and sends back the response. Once signatures from majority of the members have been aggregated, the final response message $\lbrace\mathcal{I}, \mathcal{H}(\mathcal{I}), \mathbb{B},[\mathcal{H}(\mathcal{I})]_{\mathbb{S}_{\mathbb{B}}}\rbrace$ is posted in the public blockchain. The authenticity of this message can be easily verified using the public keys of the members who have signed the message, and the integrity can be checked by computing and comparing the hash of $\mathcal{I}$. This verification process is carried out by the \textit{Consumer Client} and is transparent to all the consumers. The \textit{Consumer Client} only accepts those messages which have the required number of signatures ($> \frac{2}{3} |consortium|$) along with the proper hash.

\textbf{Posting Private Information for a Consumer:} Posting private information to a consumer through the public blockchain requires some mechanism to preserve confidentiality. This is done by encrypting the message using the public key $\mathcal{P}_{\mathcal{U}}$ of the consumer $\mathcal{U}$. The message is also similarly authenticated using the aggregated multi-signature of the consortium members. Thus the final message which is posted in the public blockchain is $\lbrace<\mathcal{M}>_{\mathcal{P}_{\mathcal{U}}}, \mathcal{H}(<\mathcal{M}>_{\mathcal{P}_{\mathcal{U}}}), \mathbb{B},[\mathcal{H}(<\mathcal{M}>_{\mathcal{P}_{\mathcal{U}}})]_{\mathbb{S}_{\mathbb{B}}}\rbrace_{\mathcal{P}_{\mathcal{U}}}$, where $<\mathcal{M}>_{\mathcal{P}_{\mathcal{U}}}$ denotes a message $\mathcal{M}$ encrypted using the key $\mathcal{P}_{\mathcal{U}}$. Thus, only the consumer $\mathcal{U}$ can decrypt the message using its secret key $\mathcal{S}_{\mathcal{U}}$. The \textit{Consumer Client} handles the decryption and verification of authenticity.

\subsection{Optimizing the Latency for Signature Collection} 
Since the messages to be transferred from the consortium to the consumers already have to be committed in the private blockchain, the multi-signature collection process is decoupled and carried out off-chain to improve the latency. Thus the consortium members communicate through peer-to-peer messages to form the verifiable signed message. This multi-signature mechanism's latency depends on the way the members forward the messages and collect back the signatures to generate the final payload by aggregating them. Thus a communication tree is formed along which the singing request and the signatures are exchanged. One extreme case of this is when one of the members acts as the leader, and the other members sign their messages and forward them back to it. The leader constructs the collective signature by including its own signature and validates other members' signatures against their public keys.

This strategy is likely to have low latency because of its star topology with a path length of at most one but will have high signature combination computation overhead for the leader. Another extreme is to consider a linear chain of consortium members through which the above round of messages propagate; this will have less computation overhead for each member but will have high network latency. 
\our{} uses a $M$-ary tree structure to propagate multi-signature collection messages through which individual signatures are collected, and the multi-signature is constructed following Eq. (\ref{eq:bls}). 
Interestingly, the latency for multi-signature generation changes with the value of $M$, which we analyze in Section~\ref{sec:eval}.  

\textbf{Handling denial of service:} Off-chain multi-signature collection improves the latency of the process. However, it introduces the risk of denial of service. Although the message to be signed is first committed in the private blockchain through the consensus process, some malicious consortium participants may try to halt the consortium through denial of service attack by not responding to signature collection requests. As a result, to prevent that and detect the faulty members to hold them responsible, \our{} resorts to a blockchain contract-based signature collection after the off-chain protocol fails (possibly with a timeout). For a message, the \textit{Signature Collection} contract is initialized in a similar way as \textit{Propagation Contract}, and gathers BLS signatures of the members. Thus any non-cooperating member is detected through this transparent process, who can be held responsible.

%
%
%


\section{Use Case Implementation: Cloud Federation}
To evaluate the potential of \our{}, we have implemented a use-case of cloud federations like \textit{OnApp}, where a group of CSPs participate in a single marketplace to offer cloud infrastructure such as virtual machines (VMs) as a service (IaaS) to the consumers. Traditionally cloud brokers~\cite{brokerprofittpds} or centralized marketplaces like \textit{OnApp} coordinate all interactions between the CSPs and the consumers. To design a fully trustless decentralized architecture for cloud federations, we use {\our} to implement a private network of CSPs and a public network of consumers, called \textit{CollabCloud}. Apart from the basic functionalities of {\our}, \textit{CollabCloud} implements a {Fair Scheduling Contract} within the CSP consortium to schedule the VM requests among the participating CSPs while ensuring fairness in terms of profitability of the CSPs and quality of service (QoS) for the consumers.

\label{appen:fairschedule}
\begin{algorithm}[!t]
    \scriptsize
	\SetAlgoLined
	\KwIn{$\mathcal{R}_{i}$, $\mathbb{K}$, $\mathbb{W}$ }
	\KwResult{Scheduled CSP: $\mathcal{C}_s$ }
	\For{ $\mathcal{C}_{j} \in \mathcal{F}$}{
		\textbackslash* Initialize current proportion of scheduled requests of $\mathcal{C}_{j}$ to 0 *\textbackslash \\
		$\mathcal{G}_{\mathcal{C}_{j}} \leftarrow 0$\\
		\For{$l \leftarrow 1~ to~ |\mathbb{W}|$}
		{
			\If{$\mathbb{W}[l] = \mathcal{C}_{j}$}
			{
				$\mathcal{G}_{\mathcal{C}_{j}} \leftarrow \mathcal{G}_{\mathcal{C}_{j}} + 1$
			}
		}
		$\mathcal{G}_{\mathcal{C}_{j}} \leftarrow \frac{\mathcal{G}_{\mathcal{C}_{j}}}{|\mathbb{W}|} $\\
		$\mathcal{D}_{\mathcal{C}_{j}} \leftarrow \mathcal{G}_{\mathcal{C}_{j}} - \hat{\mathcal{K}_{\mathcal{C}_{i}}}$
		
	}
	$\mathcal{C}_{s} \leftarrow \operatorname*{argmax}_{\mathcal{C}_{i} \in \mathcal{F}} (\mathcal{D}_{\mathcal{C}_{j}})$
	\caption{Fair Request Scheduling Contract}
	\label{algo:scheduling}
	enqueue($\mathbb{W}$, $\mathcal{C}_{s}$)\\
	\If{$|\mathbb{W}| > \mathfrak{K}$}{	dequeue($\mathbb{W}$)\\
	}
	\textbf{return} $\mathcal{C}_{s}$
\end{algorithm}

The \textit{Fair Request Scheduling Contract} takes into account the contribution of the individual CSPs in the federation and schedules consumer requests in proportion to it. We define the federation $\mathcal{F} = \{ \mathcal{C}_{1}, \mathcal{C}_{2}, \ldots, \mathcal{C}_{n} \}$ as a collection of CSPs $\mathcal{C}_{i}$. A CSP $\mathcal{C}_i$ can support certain VM configurations which are represented by $\mathbb{V}^{\mathcal{C}_i} = \{ \mathcal{V}_1,\mathcal{V}_2, \ldots, \mathcal{V}_m \}$. Thus the catalog of the federation is the union of all such VM configurations being offered by the individual CSPs, represented as $\mathbb{C} = \bigcup\limits_{\mathcal{C}_i \in \mathcal{F}}  \mathbb{V}^{\mathcal{C}_i}$.  Similar to the catalog, the contribution of each CSP $\mathcal{C}_i$ is a set of \textit{VM offerings}, denoted by $\mathbb{O}^{\mathcal{C}_{i}} = \{ \mathcal{O}_1, \mathcal{O}_2, \ldots,\mathcal{O}_m \}$. A \textit{VM offering} is defined as a three-tuple: $\mathcal{O} = \{\mathcal{V},k,c\}$, where $\mathcal{V}$ denotes a VM configuration, $k$ denotes the quantity of the VMs of the particular configuration the CSP can offer, and $c$ denotes the expected pricing of that VM type. A consumer request for a VM is defined as a four-tuple: $\mathcal{R} = \{ \mathcal{R}_{id}, \mathcal{P}_{\mathcal{U}},  \mathcal{V}_{j}, \mathcal{D} \}$, where $\mathcal{R}_{id}$ is the unique identifier of the consumer request, $\mathcal{P}_{\mathcal{U}}$ is the public key of the consumer making the request, $\mathcal{V}_{j} \in \mathbb{C}$ is the VM configuration selected from the catalog $\mathbb{C}$, and $\mathcal{D}$ is the duration for which the VM is requested.

The fair scheduling smart contract is shown in Algorithm~\ref{algo:scheduling}. The input to the algorithm is a consumer request $\mathcal{R}_i$, the proportions of contribution of all CSPs in the federation $\mathbb{K} = \{\hat{\mathcal{K}_{\mathcal{C}_{i}}} | ~\mathcal{C}_{i} \in \mathcal{F} \}$, and an array $\mathbb{W}$ consisting of the results of this algorithm for last $|\mathbb{W}|$ scheduled requests. We define infrastructure contribution $\mathcal{K}_{\mathcal{C}_{i}}$  of each CSP $\mathcal{C}_i$ as
$
\tiny
\mathcal{K}_{\mathcal{C}_{i}} = \sum_{\mathcal{O} \in \mathbb{O}^{\mathcal{C}_{i}}} \mathcal{O}.\mathcal{V}.\mathcal{CPU} \times \mathcal{O}.k
$, that is the weighted sum of the quantities of its \textit{VM offerings} indicating the amount of IaaS capacity (hardware resources) contributed. Thus, each time the catalog is updated, the proportion of contributions are also changed. The contribution proportion is thus $\tiny
\hat{\mathcal{K}_{\mathcal{C}_{i}}} = \frac{\mathcal{K}_{\mathcal{C}_{i}}} {\sum_{\mathcal{C}_{j} \in \mathcal{F}} \mathcal{K}_{\mathcal{C}_{j}}} 
$, for each CSP $\mathcal{C}_{i}$.

In essence, the scheduler works similarly to a \textit{weighted fair queue}~\cite{wfq} which ensures that the rate of consumer requests received by each CSP $\mathcal{C}_i$ is proportional to $\hat{\mathcal{K}_{\mathcal{C}_{i}}}$.
For this purpose, the scheduling contract keeps track of a window ($\mathbb{W}$) of the past scheduled results. We implement $\mathbb{W}$ as a queue containing results for past requests that is $\mathcal{R}_{i-1}, \mathcal{R}_{i-2}, \ldots, \mathcal{R}_{i-|\mathbb{W}|}$. Here each result corresponds to some CSP to which the past request was scheduled. The algorithm first computes the proportion of requests scheduled to a particular CSP as $\mathcal{G}_{\mathcal{C}_{j}}$ and then computes the proportion deficit as $\mathcal{D}_{\mathcal{C}_{j}}$. The request $\mathcal{R}_{i}$ is then scheduled to the CSP, having the maximum deficit in its share of past scheduled requests. Then the window $\mathbb{W}$ is updated by inserting the new result, and also removing the oldest result if $|\mathbb{W}| > some~threshold~\mathfrak{K}$.

\textbf{Verifiability of the Scheduling Algorithm:} $\mathcal{R}_{i}$ is obtained from the public blockchain, and $\mathbb{K}$ is available in the private ledger. Finally, the past scheduled requests are obtained from the previous results of the \textit{Fair Resource Scheduling} contract in the private blockchain. Thus, each CSP has access to all the information from the two blockchains. For verifiability, it must be ensured that all the CSPs act on the same version of information. With each execution of \textit{Fair Resource Scheduling} contract, the value of $\mathbb{W}$ is altered. Therefore, the CSPs must know which version of $\mathbb{W}$ is applicable for which transaction. This is ensured in two different ways -- \textbf{order-execute} and \textbf{execute-order} based executions of the contracts~\cite{fabric}.
	
\noindent\textbf{Case (i): order-execute -- } The transactions are ordered first, and the consensus is achieved on this ordering. The transactions are then executed sequentially based on the agreed order, and $\mathbb{W}$ is updated. Thus every CSP applies the transactions in the same sequence on $\mathbb{W}$, starting from the initial version.
	
\noindent\textbf{Case (ii): execute-order --} Each transaction is first simulated on a particular version of $\mathbb{W}$, and this version number is also included in the transaction. Then the simulation result (an updated version of $\mathbb{W}$) is sent for consensus. In the case of multiple such parallel transactions acting on the same version of $\mathbb{W}$, only one transaction is agreed upon during the consensus and accepted. The rest of them are rejected.

\begin{figure}[!ht]
	\centering 
	\includegraphics[width=0.85\linewidth]{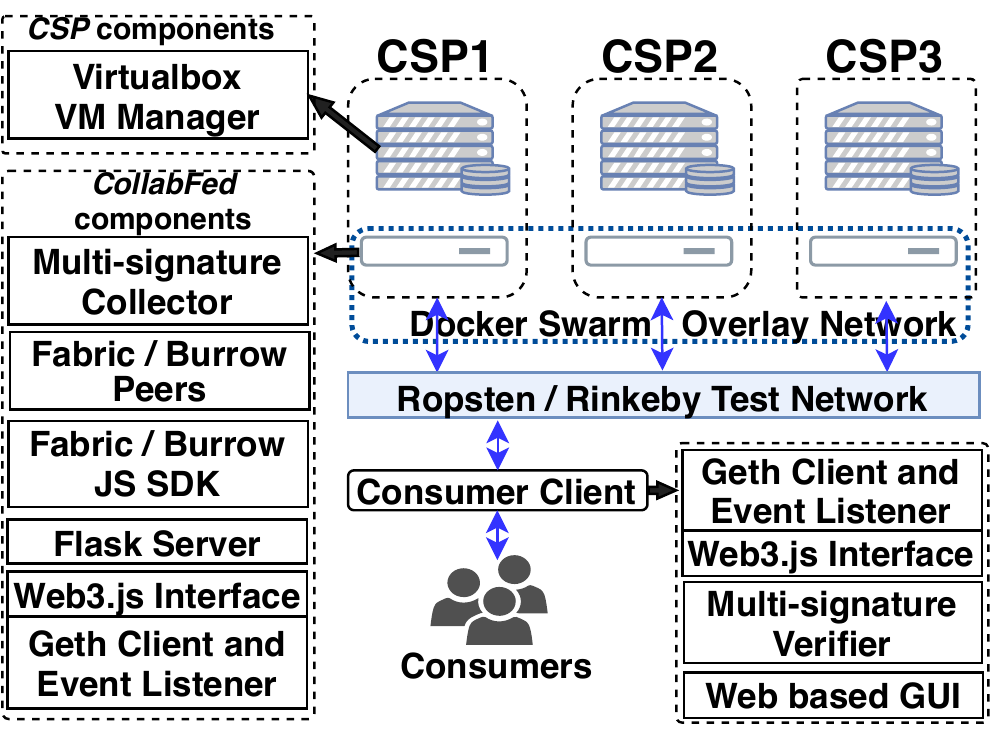}
	\caption{\textit{CollabCloud} modules and Testbed setup}
	\label{fig:testbed}
\end{figure}

\section{Evaluation}
\label{sec:eval}
In order to test the feasibility and practicality of \our{}, we have implemented a PoC of \textit{CollabCloud} decentralized cloud federation. Each component of \our{} along with the additional cloud federation specific functionalities are developed, and the end-to-end system is deployed in a testbed (Fig.~\ref{fig:testbed}). Since \our{} needs one public blockchain platform for providing the \textit{Consortium Interface}, we have chosen \textit{Ethereum}~\cite{ethereum}. For the private blockchain, we have tested with \textit{Hyperledger Fabric} and \textit{Burrow} platforms. The public blockchain smart contracts are implemented using \texttt{Solidity (v0.5.0)} (\url{https://solidity.readthedocs.io/en/v0.5.0/}) language, and they are executed on the Ethereum Virtual Machine (EVM). We have used \texttt{Truffle} (\url{https://www.trufflesuite.com/}) for the development and testing of the Ethereum contracts. Two test networks (\url{https://docs.ethhub.io/using-ethereum/test-networks/}), \texttt{Ropsten} and \texttt{Rinkeby} are used to run the consortium interface. \texttt{Ropsten} uses Proof of Work (PoW) whereas \texttt{Rinkeby} uses Proof of Authority (PoA)~\cite{de2018pbft} for consensus. We evaluate \our{}, as well as the cloud-federation functionalities from two different setups. First, we develop an in-house testbed with three emulated CSPs over six cloud servers (each CSP having two servers). Next, to analyze the scalability of different components of \our{}, we perform an emulation-based evaluation over the Mininet virtual emulation network~\cite{mininet}. 

\subsection{Platform Setup}
\label{sec:testbedsetup}
To test the end-to-end functionality and performance of \our{} along with its various components, we set up a PoC testbed of cloud federation emulating $3$ CSPs participating in the federation. Fig.~\ref{fig:testbed} shows the setup where each CSP has two cloud servers -- one $4$-core Intel Core i5-4590@3.30GHz server with $8$GB memory (Ubuntu 18.04, Linux Kernel 4.15) for running \our{} services, and another $88$-core Intel Xeon Gold 6152@2.10GHz server with $256$GB memory (CentOS 7.7, Linux Kernel 3.10) for running the CSP's usual services including VM placement and hosting the VMs. All the services are run in \texttt{Docker} (\url{https://www.docker.com/}) containers, and the networking is established through a Docker swarm overlay network. 

For implementing the CPS functionalities, we have used \texttt{VirtualBox} (\url{https://www.virtualbox.org/}) for creating VMs, and a \texttt{Flask} (\url{https://flask.palletsprojects.com/en/1.1.x/}) server for accepting VM placement requests and interfacing with \texttt{VirtualBox}. Since each CSP has only one emulated data center, which is the host server itself, the placement algorithm does not affect our system's evaluation. However, each CSP has its own set of supported VM specifications that it offers, resulting in different catalogs. We use the \textit{Fair Request Scheduling} contract that allocates requests based on the proportionality of the virtual CPU (vCPU) contribution in the federation by each CSP.  

Apart from the testbed, to evaluate the scalability of different components of \our{}, we also created a Mininet-based network topology for emulation. We created test scenarios with several \textit{CollabCloud} CSP nodes ranging from $2$ to $32$ and the latency between them ranging from $50$ms to $400$ms to capture their performance in real-world deployments.

\subsection{End-to-end Testbed experiments}
In these experiments, we used the PoC testbed to evaluate each component's performance while doing end-to-end consumer request processing for VM provisioning. We used emulated consumers with numbers ranging from $4$ to $64$ and programmed them to send parallel requests at the same instance of time. We have evaluated the latency and overheads of processing these requests in each \our{} module.

\begin{figure}[!ht]
	\begin{minipage}[b]{0.49\linewidth}
	\centering
	\includegraphics[width=1\linewidth]{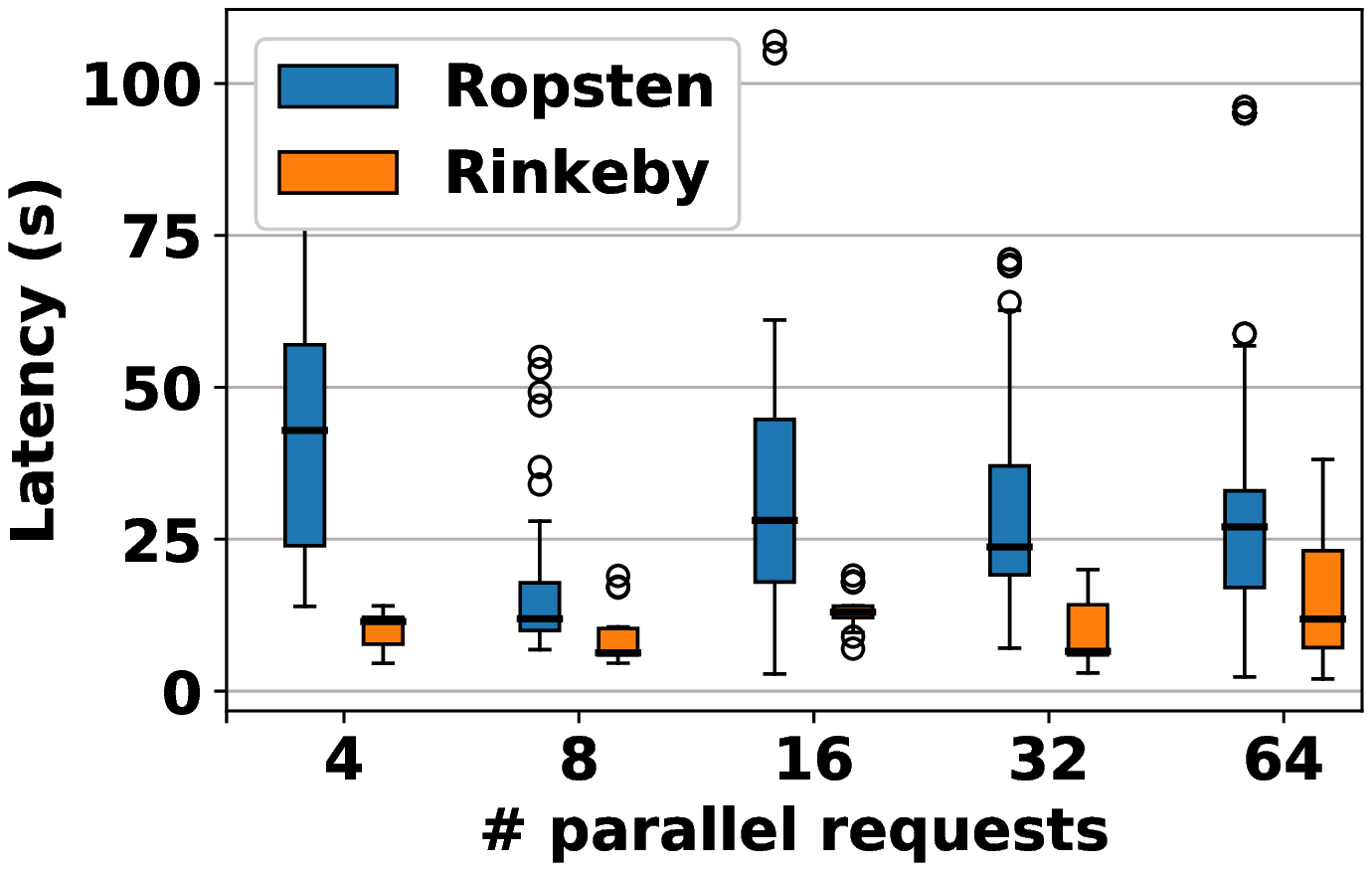}
	\caption{Public Blockchain Latency}
	\label{fig:public_blockchain_latency}
	\end{minipage}
	\hspace{0.2pt}
	\begin{minipage}[b]{0.49\linewidth}
	\centering 
	\includegraphics[width=1\linewidth]{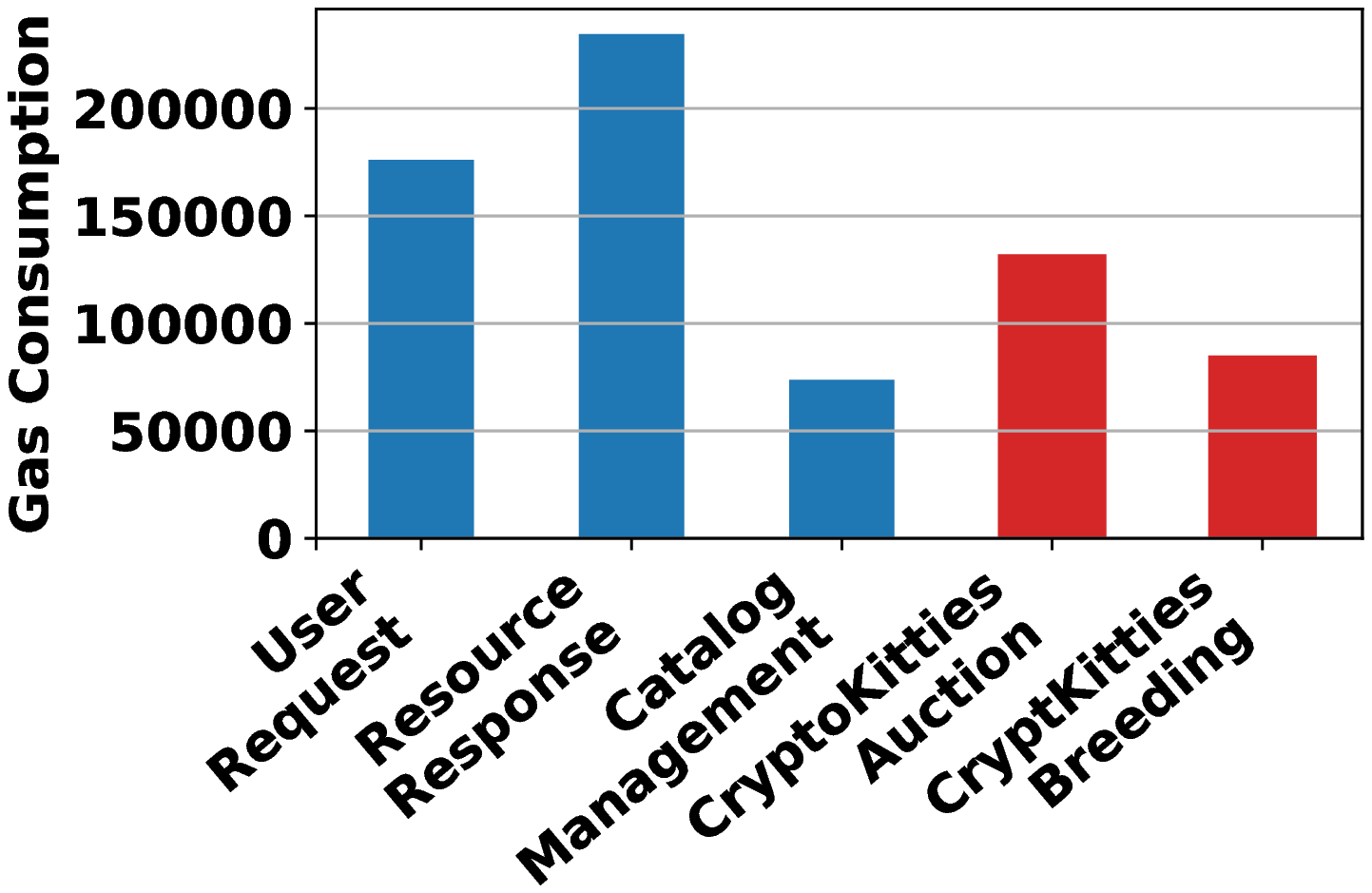}
	\caption{Gas Consumption}
	\label{fig:gas}
	\end{minipage}
\end{figure}
\textbf{Consortium Interface}: Each consumer request encounters the public blockchain twice, first when it propagates from the public blockchain to the consortium, and then in the \textit{Resource Response Contract}, when the processed result is transferred back from the private blockchain to the public one. Fig.~\ref{fig:public_blockchain_latency} shows the distribution of latency for processing the consumer requests over the public Ethereum blockchain. The processing latency over Ethereum test networks varies widely at different times depending upon the usage by other Ethereum users across the globe. We have collected the data for two weeks at different times of the day, and the same has been plotted in Fig.~\ref{fig:public_blockchain_latency}. We observe that the PoW-based consensus process of Ropsten test network has a higher transaction processing time compared to the PoA-based Rinkeby network.

%

Each contract in the public blockchain requires some transaction fees proportional to its computational complexity or storage requirements. In Ethereum, this is measured as ``Gas". Fig.~\ref{fig:gas} shows the gas consumption of the smart contracts of \our{}, along with the cloud federation specific contracts. We observe that the \textit{Resource Provisioning} contract is of the highest complexity since it has to store the multi-signatures for each transaction and the encrypted resource access information. To understand whether this Gas requirement is too high or too low, we benchmark these values concerning the Gas consumption by \texttt{CryptoKitties} ({https://www.cryptokitties.co/}) which is a common Ethereum application, and found that they are comparable.


\begin{figure}[ht]
	\begin{minipage}[b]{0.49\linewidth}
	\centering
	\includegraphics[width=1\linewidth]{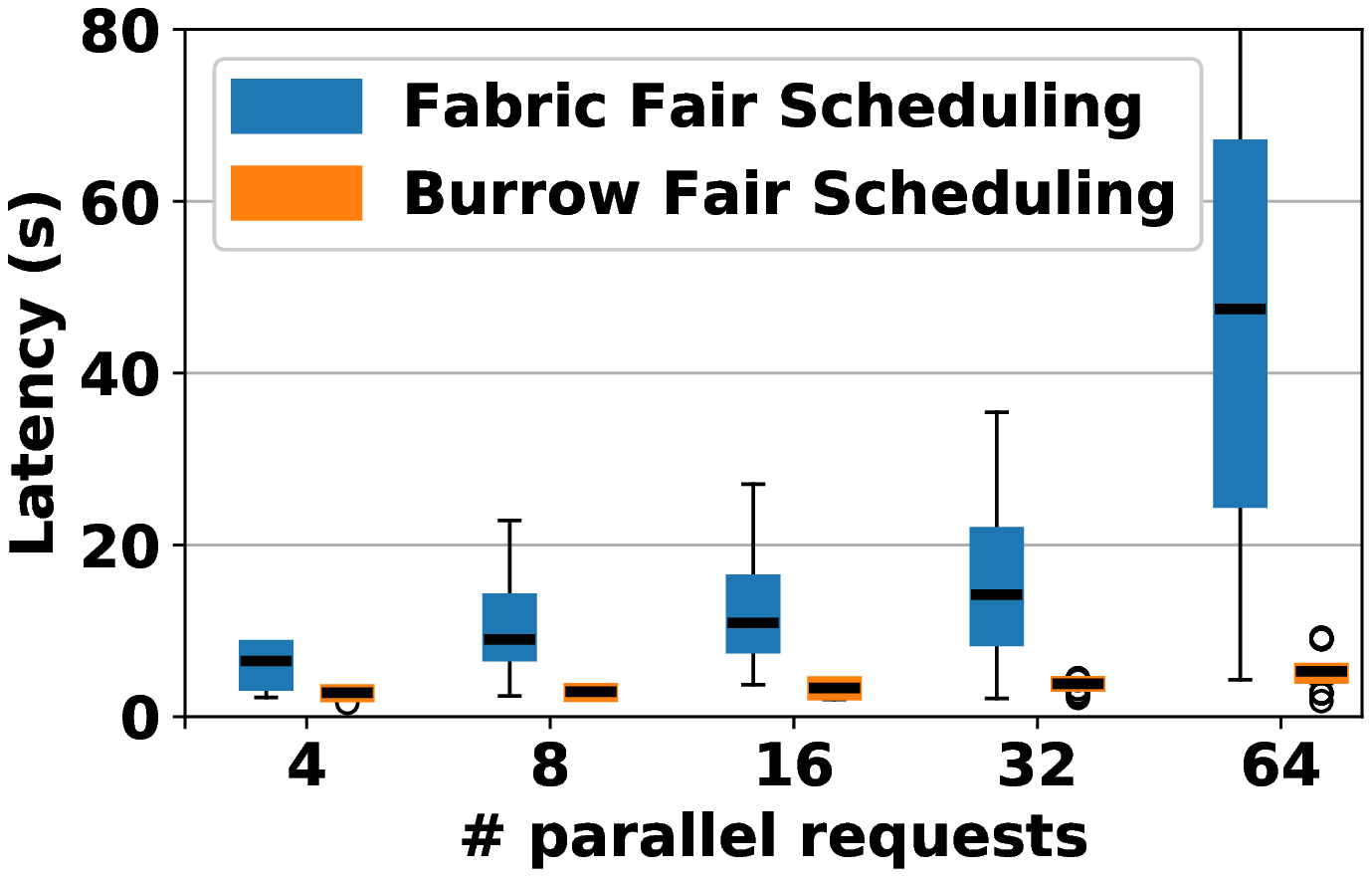}
	\caption{\textit{Fair Scheduling} Latency: Fabric vs Burrow}
	\label{fig:fabric_vs_burrow}
	\end{minipage}
	\hspace{0.2pt}
	\begin{minipage}[b]{0.49\linewidth}
		\centering
		\includegraphics[width=1\linewidth]{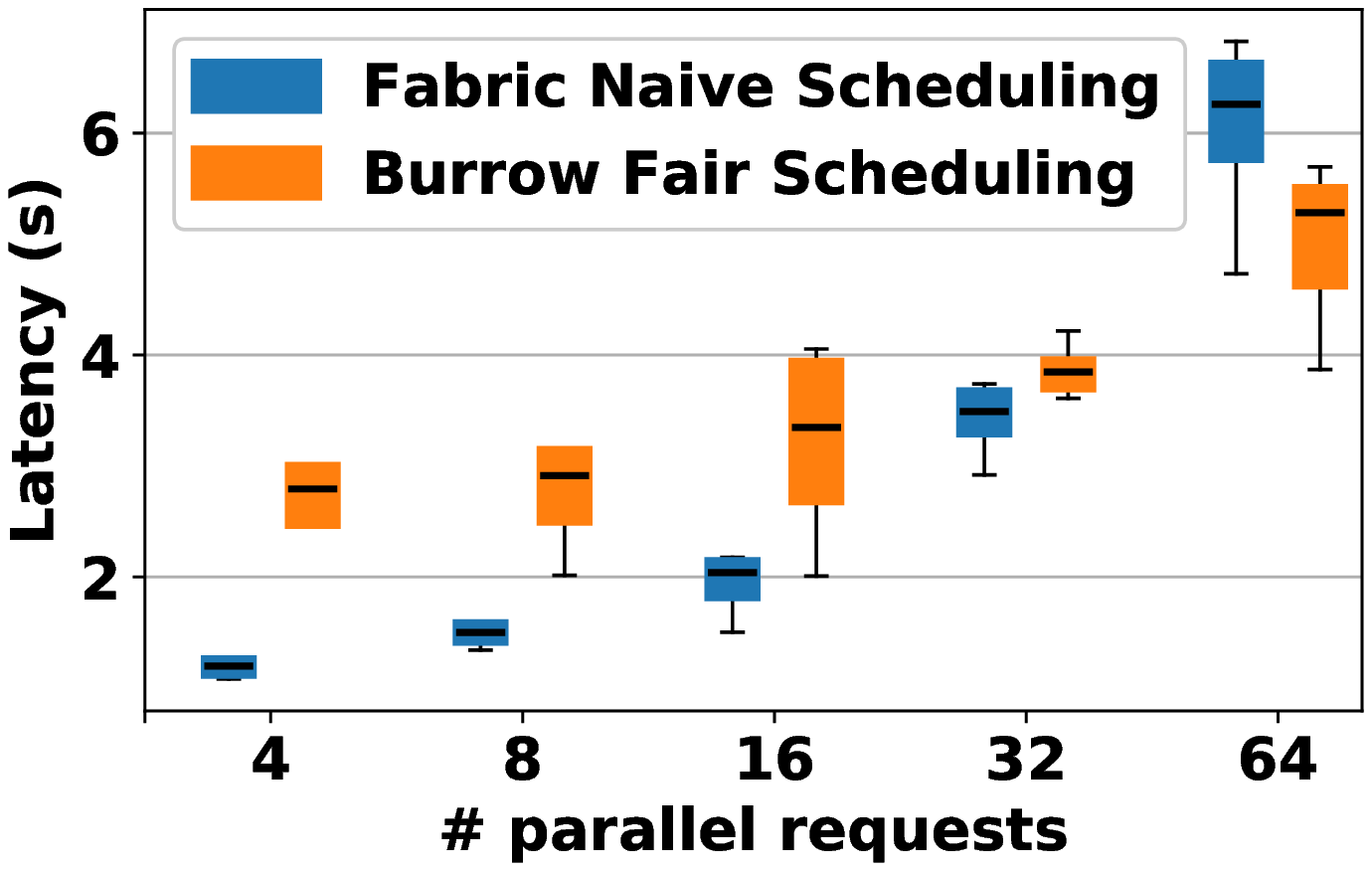}
		\caption{Fabric Static Scheduling vs Burrow Fair Scheduling}
		\label{fig:naivefabric_vs_burrow}
	\end{minipage}
\end{figure}

\textbf{Request Scheduling over Private Ledgers}: Fig.~\ref{fig:fabric_vs_burrow} shows the time required for executing fair scheduling contracts in the private blockchain. We observe that the transaction processing time for Fabric is much higher than that of Burrow. The reason for this result is specific to the type of processing required by the \textit{Fair Request Scheduling}. The key difference between Burrow and Fabric is the transaction execution workflow followed by them. Fabric follows \textit{execute-order} flow, while Burrow follows \textit{order-execute}. Executing first and then committing the results introduces a new problem for the type of contracts that read and change the system's common state, just like the \textit{Fair Request Scheduling} uses a history of the already scheduled requests at different CSPs. The reason is as follows. While executing multiple transactions in parallel, let's assume that they get executed on the same current state $\mathcal{S}_c$, and thus the output is based on $\mathcal{S}_c$. After that, once any one of the transactions is committed, the current state is changed to $\mathcal{S}'_c$. This state change also might change the output of other transactions that would be executed after it. As a result, when the other parallelly executed transactions are processed for committing, they fail in the ordering and validation phase since their execution results do not match with the execution result on $\mathcal{S}'_c$. Fabric does not retry to execute the failed transactions by itself, so \our{} over Fabric reschedules the failed transactions, thus increasing the latency.


To validate our hypothesis regarding the source of higher overhead caused by Fabric, we also tested with a naive scheduling contract that schedules the requests based on a static rule depending on its ID. This contract does not depend on the current state of the blockchain. In Figure \ref{fig:naivefabric_vs_burrow}, we can see that the scheduling latency of Fabric has dramatically improved. We also noticed that there are no transaction failures due to inconsistent execution results. Moreover, we saw no such latency improvements for Burrow with such a naive scheduling contract. It may be noted that for more parallel requests, Burrow still performs marginally better than Fabric. Consequently, we can conclude that the choice of private blockchain technology depends heavily on the fair scheduling contract's business logic.  

\begin{figure}[ht]
	\begin{minipage}[b]{0.49\linewidth}
		\centering
		\includegraphics[width=\linewidth]{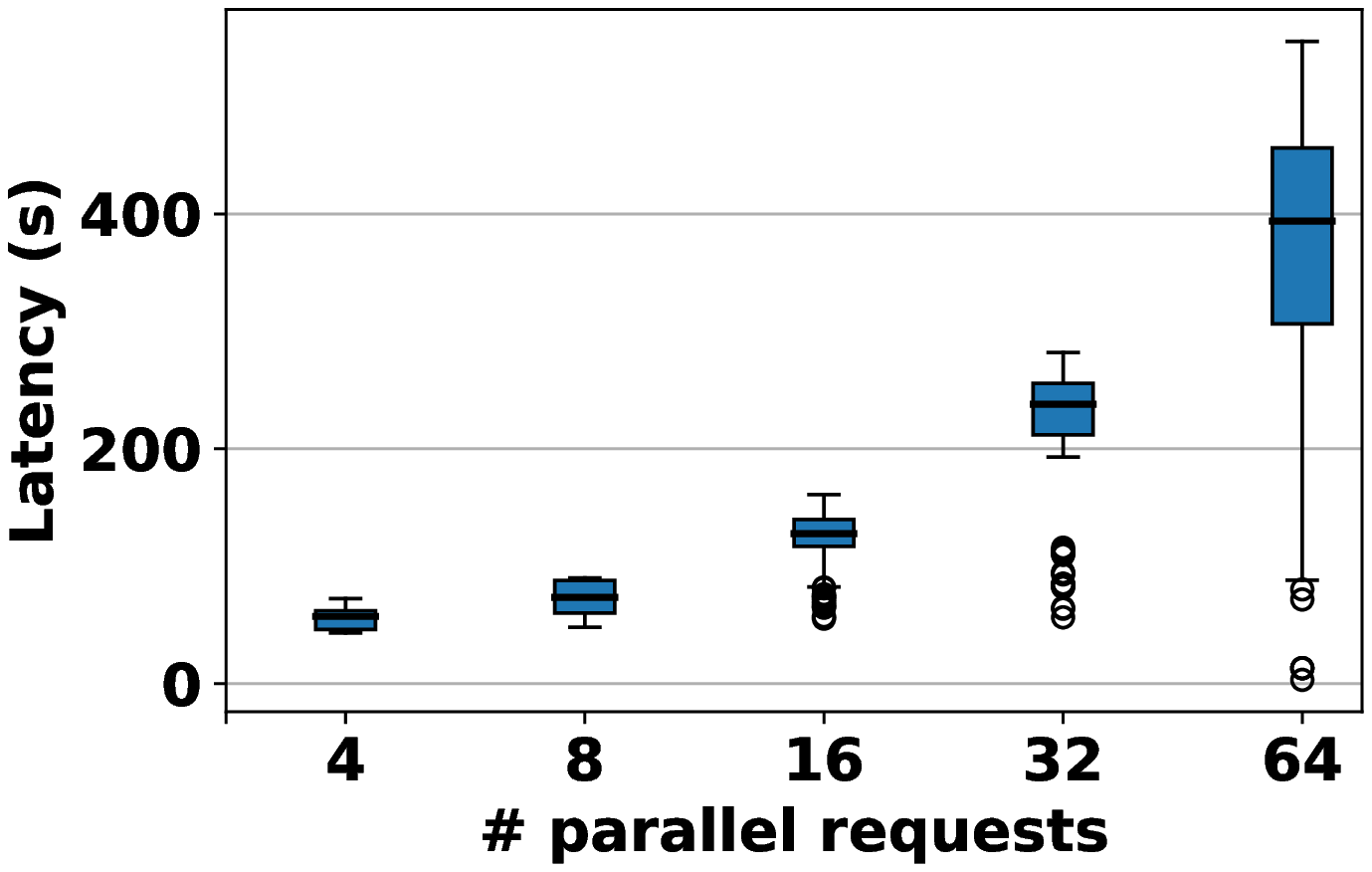}
		\caption{VM Provisioning Latency }
		\label{fig:vmprovision}
		
	\end{minipage}
	\hspace{0.2pt}
	\begin{minipage}[b]{0.49\linewidth}
		\centering
		\includegraphics[width=\linewidth]{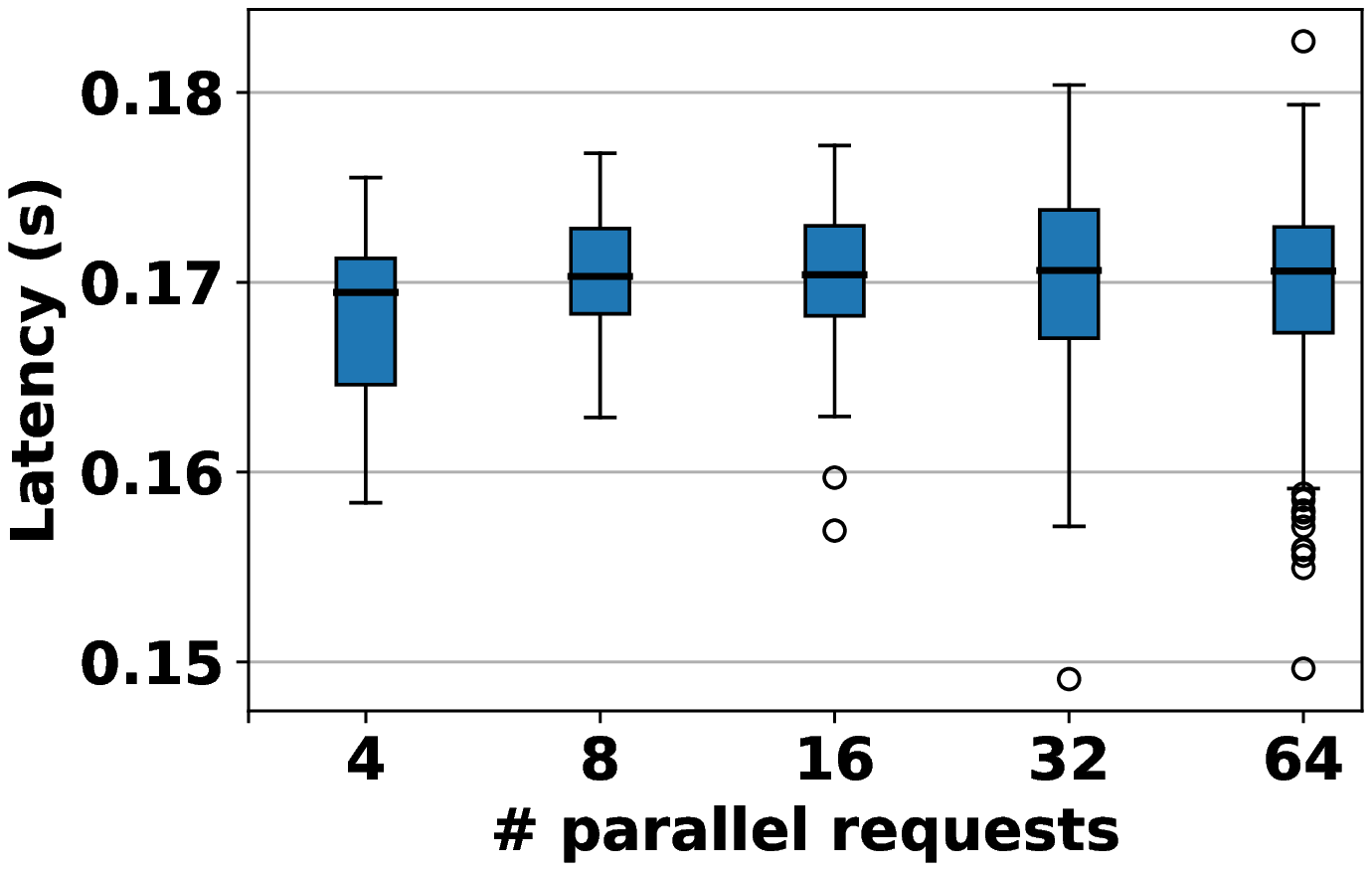}
		\caption{Sign. Collection Latency }
		\label{fig:multisignature}
	\end{minipage}
\end{figure}
%
%
After a request is scheduled, a VM is provisioned accordingly, and the access information is signed through collections of BLS multi-signatures. Fig.~\ref{fig:vmprovision} shows the distribution of the time taken for VM Provisioning. This increases with the increase in number of parallel requests, mainly due to the limited processing capability of the hardware of our setup. This latency is specific to the cloud federation application of \our{}, and thus does not count towards its performance. Fig.~\ref{fig:multisignature} shows the distribution of latency for multi-signature collection. We see that the multi-signature collection latency remains fairly consistent.

\begin{figure}[ht]
	\begin{minipage}[b]{0.49\linewidth}
		\centering
		\includegraphics[width=\linewidth]{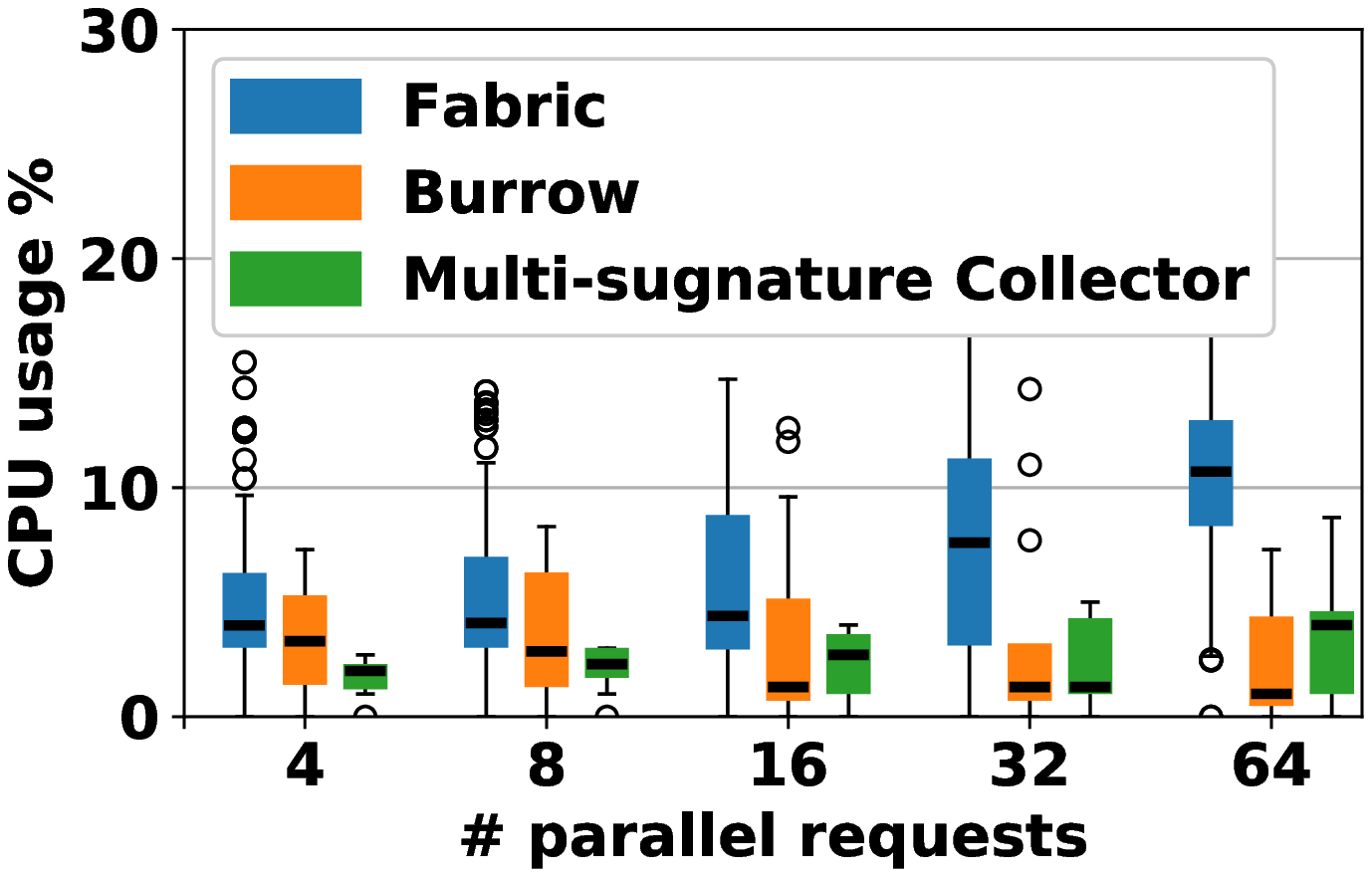}
		\caption{CPU Usage}
		\label{fig:cpu}
	\end{minipage}
	\hspace{0.2pt}
	\begin{minipage}[b]{0.49\linewidth}
		\centering
		\includegraphics[width=\linewidth]{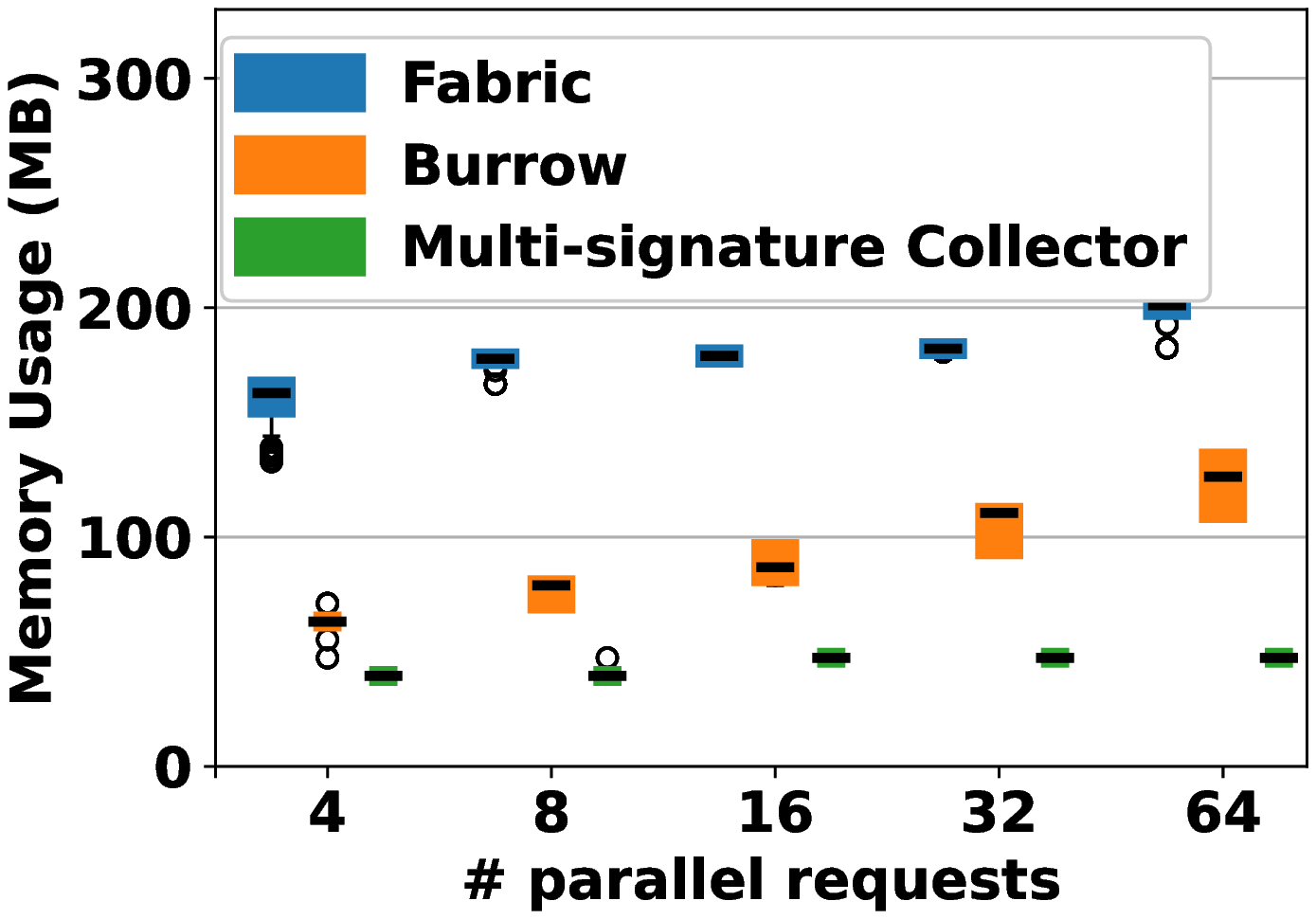}
		\caption{Memory Usage}
		\label{fig:memory}
	\end{minipage}
\end{figure}

\textbf{Resource Consumption:} {\our} consumes CPU, memory, and network bandwidth, which are an additional overhead to normal operations of a consortium. Fig.~\ref{fig:cpu} shows the box-plot distribution of CPU usage by \our{} server for executing the private blockchain transactions in Fabric and Burrow, and for multi-signature collection. We observe that the CPU consumption is reasonably low, below $10\%$  in most cases for all the services. Similarly, Fig.~\ref{fig:memory} depicts the distribution for memory requirements which stays below $200$MB. 

%

\subsection{Mininet scalability experiments}
The public blockchain platforms being open networks have been designed to be scalable, and extensive research has been done to study their performance~\cite{byzcoin,algorand}. We focus on the scalability of the private network and the multi-signature collection. For this, we set up an experiment with $32$ emulated CSPs over a Mininet~\cite{mininet} topology, which forms a \our{} consortium. We also changed the inter-CSP network latency to emulate the CSPs' spread across different geographic regions.


\begin{figure}[ht]
	\begin{minipage}[b]{0.49\linewidth}
		\centering 
	\includegraphics[width=\linewidth]{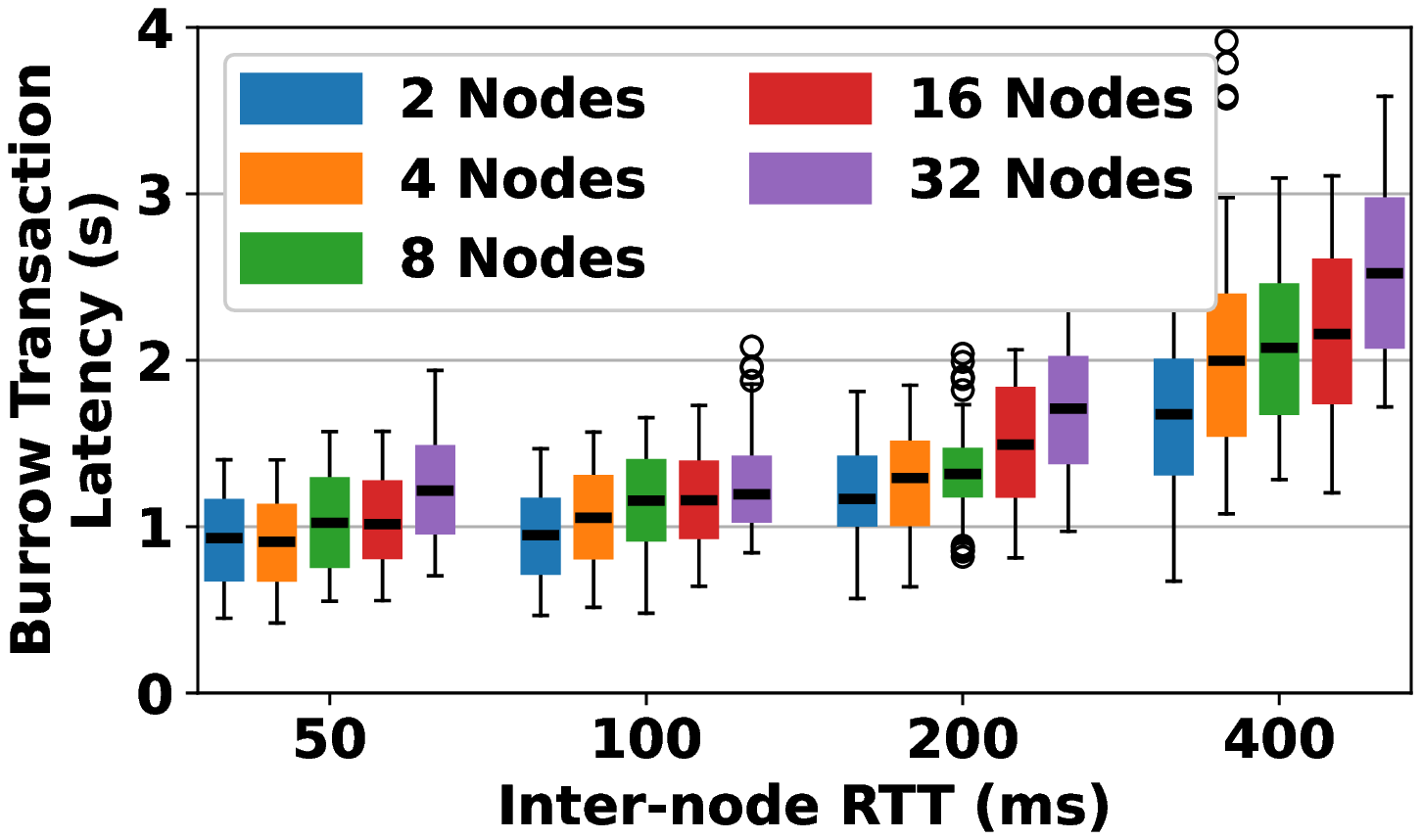}
	\caption{Burrow scalability}
	\label{fig:burrowscalability}
	\end{minipage}
	\begin{minipage}[b]{0.5\linewidth}
		\centering 
	\includegraphics[width=\linewidth]{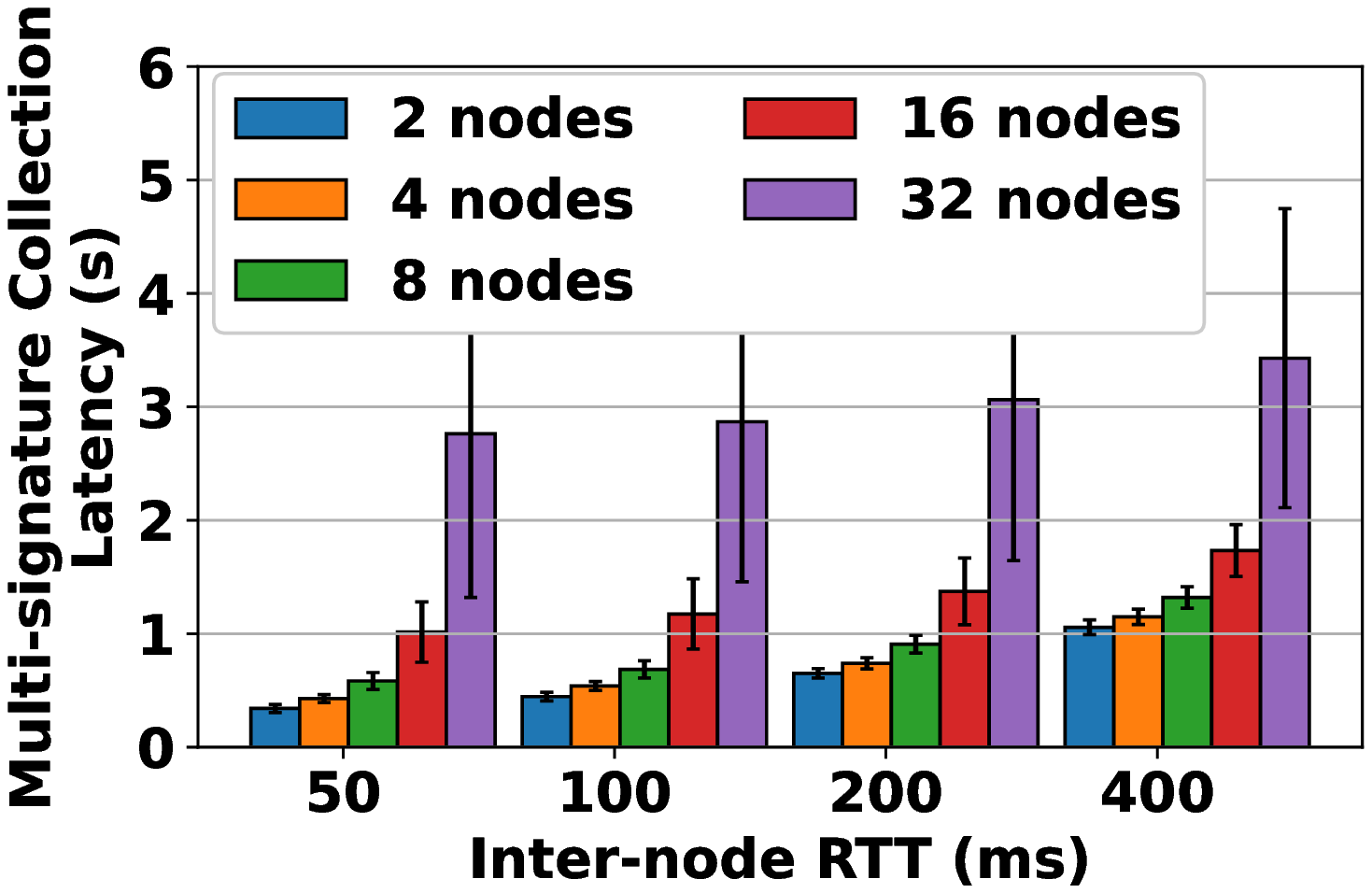}
	\caption{BLS scalability}
	\label{fig:multisignature_scalability}
	\end{minipage}
\end{figure}

Fig.~\ref{fig:burrowscalability} shows the distribution of Burrow propagation contract execution and commitment latency. The experiment has been done with inter-CSP latency varying in each case, from 50ms to 400ms. We observe that the median transaction latency lies around $2.5$ seconds with 32 nodes and 400ms inter-CSP latency. Further, the increment in the transaction latency due to an increase in the number of CSPs or inter-CSP network latency is not very high, which indicates the scalability of the proposed approach.

%

Fig.~\ref{fig:multisignature_scalability} presents the mean and the standard deviation of multi-signature aggregation latency with a varying number of nodes and inter-CSP latency values. We observe that the mean latency is below $2$ seconds for $16$ SPs and about $3.5$ seconds for $32$ SPs. This also indicates the scalability of the signature aggregation scheme. However, the multi-signature collection latency can have a big impact due to the collection tree structure. To study it, we constructed a complete M-ary communication tree with $32$ SPs. Table~\ref{table:multisignature} shows the multi-signature collection latency for different values of $M$. The inter-CSP latency for this test is kept fixed at $400$ms. We can observe a sharp improvement in the latency from linear (M=1) to binary tree (M=2) structure. The latency is more or less stable from $M=4$. However, the multi-signature combination complexity for individual CSPs increase with the value of $M$. Therefore, the value of $M$ in a real deployment can be chosen based on this trade-off. 

\begin{table}[!ht]
	\scriptsize
	\centering
	\caption{Effect of communication tree on multisig collection latency}
	\label{table:multisignature}
	\begin{tabular}{|c|c|c|c|c|c|c|c|}
		\hline
		M                                                             & 1    & 2   & 4   & 6   & 8   & 16  & 31  \\ \hline
		\begin{tabular}[c]{@{}c@{}}Mean Latency (s)\end{tabular}   & 29.9 & 5.0 & 3.2 & 2.3 & 2.4 & 3.0 & 2.9 \\ \hline
		\begin{tabular}[c]{@{}c@{}}Standard Deviation\end{tabular} & 1.8  & 0.3 & 0.2 & 0.1 & 0.2 & 0.6 & 1.3 \\ \hline
	\end{tabular}
\end{table}
\section{Conclusion}
Towards a fully trustless decentralized architecture for an electronic business consortium providing services to consumers, {\our} introduces a public-private hybrid blockchain architecture with a unified interface between the consortia and the open network. To the best of our knowledge, this is the first attempt to fill a critical gap in the application of blockchain in the enterprise and business use cases. {\our} is flexible in terms of the choice of public and private blockchain networks; however, the performance and security guarantees depend on the assumptions of those underlying blockchain technologies and consensus protocols. The PoC implementation of {\our} indicates that the system is scalable and performant with Hyperledger and Ethereum -- one of the most popular private and public blockchain platforms, respectively. The analysis of the impact of different blockchain protocols on the architecture is an exciting direction for our future works to develop a more robust system.

\bibliographystyle{IEEEtran}
\bibliography{biblio1}
\end{document}